\documentclass[aps,prl,onecolumn]{revtex4}  
\usepackage{amssymb}
\usepackage{bm}
\usepackage{amsbsy}
\usepackage{epsfig}
\usepackage{color}
\usepackage{amsmath}

\begin{document}

\date{\today}

\title{ \begin{center}
Poloidal rotation driven by nonlinear momentum transport in strong electrostatic turbulence
\end{center}}

\author{Lu Wang,$^{1*}$ Tiliang Wen,$^1$ and P. H. Diamond$^{2}$\\
$^1$State Key Laboratory of Advanced Electromagnetic Engineering and Technology, School of Electrical and Electronic Engneering, Huazhong University of Science and Technology, Wuhan 430074, China\\
$^2$Center for Momentum Transport and Flow Organization and\\Center for Astrophysics and Space Sciences,\\ University of California at San Diego, La Jolla, CA 92093-0424, USA\\
$^*$E-mail: luwang@hust.edu.cn}

\maketitle \maketitle

\section*{Abstract}
Virtually, all existing theoretical works on turbulent poloidal momentum transport are based on quasilinear theory. Nonlinear poloidal momentum flux - $\langle \tilde{v}_r \tilde{n} \tilde{v}_{\theta} \rangle$ is universally neglected. However, in the strong turbulence regime where relative fluctuation amplitude is no longer small, quasilinear theory is invalid. This is true at the all-important plasma edge. In this work, nonlinear poloidal momentum flux $ \langle \tilde{v}_r \tilde{n} \tilde{v}_{\theta} \rangle $ in strong electrostatic turbulence is calculated using Hasegawa-Mima equation, and is compared with quasilinear poloidal Reynolds stress. A novel property is that symmetry breaking in fluctuation spectrum is not necessary for a nonlinear poloidal momentum flux. This is fundamentally different from the quasilinear Reynold stress. Furthermore, the comparison implies that the poloidal rotation drive from the radial gradient of nonlinear momentum flux is comparable to that from the quasilinear Reynolds force. Nonlinear poloidal momentum transport in strong electrostatic turbulence is thus not negligible for poloidal rotation drive, and so may be significant to transport barrier formation.

\section{I. Introduction}

It is well known that poloidal rotation plays a crucial role in suppressing microturbulence through its impact on $ E \times B $ shear \cite{Biglari H. PoF 1990,Hahm T. S. PoP 1995}. The $ E \times B $ flow shear is linked to poloidal rotation via $E_r$, which is determined by the radial force balance equation, $E_r=\displaystyle \frac{\nabla P_i}{eZn} + v_{\phi}B_{\theta} - v_{\theta}B_{\phi}$. Since poloidal rotation can be significant for triggering the formation of transport barriers \cite{Zweben S J PoP 2010,Conway G D PRL 2011,Xu G S PRL 2011,Estrada T PRL 2011} through $ E \times B $ flow shear and leading to improvement of confinement and fusion performance, there have been intensive theoretical and experimental investigations into understanding of poloidal momentum transport and poloidal roation generation \cite{Diamond2005,Tynan2009} .Most theoretical works have been developed based on neoclassical calculations for both core and edge plasmas and for different collisionality \cite{R. D. Hazeltine PoF 1961,Houlberg W. A. PoP 1997,Stacey W. M. PoF 1992,W. M. Stacey PoP 2002}.

In some experiments, such as MAST\cite{A R Field PPCF 2009} and NSTX\cite{Bell R. E. PoP 2010}, the measured poloidal rotation is consistent with neoclassical predictions. This is likely due to strong neoclassical damping in spherical tokamaks. In contrast, for conventional tokamaks, such as JET \cite{K. Crombe PRL 2005}, DIII-D \cite{W. M. Solomon 2006,Chrystal2014}, found deviation of poloidal flows from neoclassical prediction for various regimes of plasmas. It is interesting to explore possible explanations for the discrepancy between neoclassical predictions and experimental observations in conventional tokamaks, which is referred as anomalous poloidal rotation \cite{Tala2007}. Turbulent drive associated with drift wave turbulence might be a promising candidate, since the anomalous transport of particle, heat and toroidal rotation is usually thought to result from drift wave turbulence. Turbulent residual stress driving intrinsic toroidal rotation has been intensively studied \cite{Diamond2013,Ida2014}.

Similar to the total flux of parallel (toroidal) momentum \cite{Diamond09}, the total poloidal momentum flux driven by electrostatic turbulence can be written as:
\begin{equation}\label{momentum flux}
\Pi_{r,\theta} = \langle n \rangle \langle \tilde{v}_r \tilde{v}_{\theta} \rangle + \langle v_{\theta} \rangle \langle \tilde{v}_{r} \tilde{n} \rangle + \langle \tilde{v}_r \tilde{n} \tilde{v}_{\theta} \rangle.
\end{equation}
Here, on the right hand side (RHS), the first term is the poloidal Reynolds stress, the second term is convection, due to the particle flux, and the last triplet term is the \emph {nonlinear} flux, which is driven by processes such as mode-mode coupling and turbulence spreading \cite{Diamond09}. There are many theoretical works on poloidal rotation driven by poloidal Reynolds stress based on quasilinear theory \cite{Diamond PoF 1991,Diamond P. H. PPCF 2008,WangL2012,C. J. McDevitt PoP 2012}. For a radial asymmetric spectrum of turbulence, a significant turbulent driven poloidal flow was predicted \cite{Diamond PoF 1991}. In \cite{Diamond P. H. PPCF 2008}, linking Reynolds force to the potential vorticity flux leads to Charney-Drazin non-acceleration theorem. Later, this work was generalized to a three-dimensional drift-ion acoustic wave system\cite{WangL2012} and to electromagnetic turbulence \cite{C. J. McDevitt PoP 2012}. Poloidal Reynolds stress driven poloidal rotation has also been observed in experiments \cite{Xu Y. H. PRL 2000,G R Tynan PPCF 2006,G. S. Xu NF 2014}. However, the \emph{nonlinear} poloidal momentum flux has been universally neglected in most existing theoretical works and simulation codes. Usually, a Boussinesq approximation implemented in fluid codes and a moment of the quasilinear flux used in gyrokinetic simulation lead to neglecting the triplet. There are no works retained the triplet and compared to the usual quadratic stress. Taking into account of Reynolds stress at the level of quasilinear theory is invalid in strong turbulence regime where the relative fluctuation amplitude is large, i.e., since $\tilde{n}/n_0 \rightarrow 1$, the nonlinear damping rate is larger than the frequency mismatch. Furthermore, recent experiments on ASDEX-U found that the triplet term could make a significant contribution to the total poloidal momentum flux for H-mode edge turbulence \cite{Muller NF 2011}. This implies that neglecting the effects of nonlinear poloidal momentum flux on poloidal rotation is not reasonable. The nonlinear parallel momentum flux is shown to be significant to intrinsic parallel rotation in strong electrostatic turbulence \cite{Wang2015}. For tokamak edge turbulence, the relative fluctuation amplitude is large, so it is possible to drive a significant nonlinear polodial momentum flux, as well. Therefore, study of the nonlinear poloidal momentum flux in the strong turbulence edge regime seems necessary for comprehensive understanding anomalous poloidal rotation and transport barrier formation physics. The rate of Reynolds work, $\langle \tilde {v}_r\tilde {v}_{\theta} \rangle^{\prime} \langle v_{\perp} \rangle$ is shown to play a key role for L-H transition on HL-2A, DIII-D and EAST experiments \cite{Tynan2013}. Its nonlinear counterpart $\langle \frac{\tilde {n}}{n_0} \tilde {v}_r\tilde {v}_{\theta} \rangle^{\prime} \langle v_{\perp} \rangle$ might be also worth investigations.

In this work, we calculate the nonlinear poloidal momentum flux using the Hasegawa-Mima (H-M) equation \cite{Hasegawa and Mima}, which is a popular drift wave model and can be reduced from Hasegawa-Wakatani model\cite{H-W} for the adiabatic electron limit. We also compare it with the quasilinear Reynolds stress presented in \cite{Diamond PoF 1991}. It is found that the turbulent poloidal rotation drive from nonlinear poloidal momentum flux is comparable to that from quasilinear Reynolds stress, particularly in steep density gradient regions. We also find that symmetry breaking in fluctuation spectrum is not required for the nonlinear poloidal momentum flux, which is fundamentally different from the Reynolds stress. Therefore, taking account of the nonlinear poloidal momentum flux effects on poloidal rotation is important in the strong turbulence edge regime. The results presented in this work indicate that turbulent Reynolds stress is incomplete for explaining the poloidal rotation in edge plasmas. Expanding the models of describing the physics of edge plasma dynamics and transport barriers formation is worthwhile.

The remainder of this paper is organized as follows. Sec.~\uppercase\expandafter{\romannumeral2} presents the minimal model and expressions for nonlinear poloidal momentum flux. We compare the nonlinear turbulent drive with the quasilinear Reynolds force in Sec.~\uppercase\expandafter{\romannumeral3}. Finally, we summarize our work and discuss the implications for anomalous poloidal rotation in Sec.~\uppercase\expandafter{\romannumeral4}. The detailed calculation is presented in the Appendix.

\section{II. Minimal theoretical model and nonlinear poloidal momentum flux}

In this section, we intend to briefly introduce the minimal theoretical model and present the expressions for nonlinear poloidal momentum flux leaving the tedious calculations in the Appendix. The interested readers can refer to our previous work \cite{Wang2015},in which nonlinear \emph{parallel} momentum flux for strong electrostatic turbulence is calculated. We will adopt a similar theoretical approach in this work.

The nonlinear poloidal momentum flux can be written as:
\begin{equation}\label{NLflux}
\Pi_{r,\theta}^{NL} =\langle \tilde{v}_r \tilde{n} \tilde{v}_{\theta} \rangle= \frac{1}{3} \left( \langle \tilde{v}_r^{(c)} \tilde{n} \tilde{v}_{\theta} \rangle + \langle \tilde{v}_r \tilde{n}^{(c)} \tilde{v}_{\theta} \rangle + \langle \tilde{v}_r \tilde{n} \tilde{v}_{\theta}^{(c)} \rangle \right).
\end{equation}
Here, $\tilde{n}$ is the density fluctuation, $\tilde{v}_r$ and $\tilde{v}_{\theta}$ are the radial and poloidal fluctuating $ E\times B $ drift velocities, respectively. The superscript (c) denotes the coherent component of the beat mode. For simplicity, we use the adiabatic approximation, i.e., $ \displaystyle \frac{\tilde{n}}{n_0} = \frac{e\tilde{\phi}}{T_e} $ and the corresponding coherent response $ \displaystyle \frac{\tilde{n}^{(c)}}{n_0} = \frac{e\tilde{\phi}^{(c)}}{T_e} $. To obtain the coherent components, $\tilde{\phi}^{(c)}$, we adopt the popular drift wave model, i.e., H-M equation\cite{Hasegawa and Mima}
\begin{eqnarray}\label{H-M}
\frac{\partial}{\partial t}\left(\rho_s^2 \nabla_{\perp}^2\phi - \phi \right) + \rho_s^4 \omega_{ci} \hat{z} \times \nabla \phi \cdot \nabla \nabla_{\perp}^2\phi -i\omega_{*n} \phi&=& 0.
\end{eqnarray}
Here, $ \hat{z} $ is the unit vector in the parallel magnetic field direction, $\nabla_{\perp}$ denotes the gradient operator perpendicular to the magnetic field direction. $\omega_{ci}= eB/(m_i c)$ is the ion gyrofrequency, $c_s$ is the ion acoustic velocity, and $\rho_s=\frac{c_s}{\omega_{ci}}$ is the ion Larmor radius at the electron temperature. We have used the standard normalization for electric potential fluctuation $\phi \equiv e \tilde{\phi} / T_e$. For the spatial scale, we consider two-scale approach, i.e., $\nabla_{\perp} = i{\bf k}_{\perp}+  {\partial}/{\partial r}$, where ${\bf k}_{\perp}$ denotes wave number of the fast spatial fluctuations, and $\partial /\partial r$ describes modulation of the wave envelope, which occurs on a slowly varying spatial scale \cite{Gurcan2006,Diamond book}. $\omega_{*n}=k_y \rho_s c_s / L_n$ is the electron diamagnetic drift frequency with $L_n = -\left(\partial \ln n /\partial r\right)^{-1}$ density gradient scale length.

Taking the Fourier transformation of Eq.~(\ref{H-M}) and neglecting higher order terms related to slow spatial variation $\displaystyle \frac{\partial ^2}{\partial r^2}$ yields
\begin{eqnarray}
\frac{\partial}{\partial t} \phi_k + i \omega_k \phi_k &= \displaystyle \sum_{k=k^{\prime} + k^{\prime\prime}} M_{k,k^{\prime},k^{\prime\prime}},\label{H-Mk}
\end{eqnarray}
where $ \omega_k =\displaystyle \frac{\omega_{*n}}{1+k_{\perp}^2{\rho_s^2}} $, and the nonlinear term is
\begin{eqnarray}
M_{k,k^{\prime},k^{\prime\prime}}&=& \displaystyle\frac{\omega_{ci}}{2 ( 1 + k_{\perp}^2 \rho_s^2 )}\rho_s^4 \ \Bigg \lbrace \hat{z}\times {\bf k}_{\perp}^{\prime} \cdot {\bf k}_{\perp}^{\prime\prime} \bigg(k_{\perp}^{\prime\prime2} - k_{\perp}^{\prime2} \bigg)\phi_{k^{\prime}}\phi_{k^{\prime\prime}}\nonumber\\
&&+ \displaystyle i\phi_{k^{\prime}} \frac{\partial}{\partial r} \phi_{k^{\prime\prime}} \bigg[k_y^{\prime} \Big(k_{\perp}^{\prime\prime2} - k_{\perp}^{\prime2} \Big) - 2k_x^{\prime\prime}\hat{z}\times {\bf k}_{\perp}^{\prime} \cdot {\bf k}_{\perp}^{\prime\prime} \bigg]\nonumber\\
&&-\displaystyle i\phi_{k^{\prime\prime}} \frac{\partial}{\partial r} \phi_{k^{\prime}} \bigg[k_y^{\prime\prime}\Big(k_{\perp}^{\prime\prime2} - k_{\perp}^{\prime2}\Big) - 2k_x^{\prime}\hat{z}\times {\bf k}_{\perp}^{\prime} \cdot {\bf k}_{\perp}^{\prime\prime}\bigg]
 \Bigg \rbrace.
\end{eqnarray}
Solving Eq.~(\ref{H-Mk}) directly, with help of the eddy-damped quasi-normal Markovian (EDQNM)\cite{Diamond book,Orszag 1970} theory, the coherent component of beat mode can be obtained\cite{Diamond book}
\begin{eqnarray}\label{coherent component}
\phi_k^{(c)}(t)= 2\int_{-\infty} ^t dt^{\prime} exp  \Big[ \big( i\omega_k + \gamma_k^{NL} \big) (t^{\prime}-t) \Big] M_{\bf k,k^{\prime},k^{\prime\prime}}.
\end{eqnarray}
Here, $\gamma_k^{NL}$ is the nonlinear damping rate, which is lager than the frequency mismatch for strong turbulence. After calculating the coherent component, the problem left is to close the forth order moment. Using the approximation of quasi-Gaussian statistics (i.e., the assumption of almost statistically independent fluctuations), \cite{Diamond book} the forth order moment can be factored into a product of quadratic moments, i.e.,
\begin{equation} \label{quadratic moments}
\langle \phi_{k^{\prime}}(t) \phi_{k^{\prime}}^*(t^{\prime}) \phi_{k^{\prime\prime}}(t) \phi_{k^{\prime\prime}}^*(t^{\prime})\rangle = \langle \phi_{k^{\prime}}(t) \phi_{k^{\prime}}^*(t^{\prime})\rangle \langle  \phi_{k^{\prime\prime}}(t) \phi_{k^{\prime\prime}}^*(t^{\prime}) \rangle.
\end{equation}
Then, with the Markovian approximation, the two-time correlation function can be expressed by one-time correlation function as
\begin{equation} \label{one-time correlation}
 \langle \phi_k^*(t^{\prime}) \phi_k(t) \rangle = exp [i\omega_k (t^{\prime}-t)-\gamma_{k}^{NL} |t^{\prime}-t| ] \langle \phi_k^*(t) \phi_k(t) \rangle.
\end{equation}
All the essentials for calculation of the nonlinear poloidal momentum flux have been obtained.

The expressions for nonlinear poloidal momentum flux are presented here directly without showing the tedious calculations. The details of calculation can be found in the Appendix. The first nonlinear flux can be written as
\begin{eqnarray}
\Pi_{r,\theta}^{NL,1}&=&\langle \tilde{v}_r^{(c)} \tilde{n} \tilde{v}_{\theta} \rangle \nonumber\\
&=& n_0 c_s^2 \Re \sum_{k=k^{\prime} + k^{\prime\prime}} ik_y \rho_s^2 \langle \phi_k^{(c)*} \phi_{k^{\prime}} ( \frac{\partial}{\partial r} \phi_{k^{\prime\prime}} + ik^{\prime\prime}_x \phi_{k^{\prime\prime}} ) \rangle  \nonumber\\
&=&\frac{1}{2}n_0 c_s^2 \sum_{k=k^{\prime} + k^{\prime\prime}}\tau_{c}\omega_{ci}\frac{I_{k^{\prime}}I_{k^{\prime\prime}}}{1+k_{\perp}^2\rho_s^2}\left( A_{k^{\prime},k^{\prime\prime}}\frac{\rho_s^2}{L_I^2}+B_{k^{\prime},k^{\prime\prime}} \right).
\end{eqnarray}
Here, $\tau_c $ is the triad interaction time for vorticity equation, and is estimated by the inverse of nonlinear damping rate, i.e., $\tau_c = \Re  \displaystyle \frac{1}{i \left( -\omega_k +\omega_{k^{\prime}}+ \omega_{k^{\prime\prime}} \right) + (\gamma^{NL}_k + \gamma^{NL}_{k^{\prime}} + \gamma^{NL}_{k^{\prime\prime}})} \cong \frac{1}{\gamma^{NL}_k + \gamma^{NL}_{k^{\prime}} + \gamma^{NL}_{k^{\prime\prime}}} $ for the reason that the nonlinear damping rate is much larger than the frequency mismatch for strong turbulence. This is opposite to the quasi-linear limit. $ I_k=|\phi_k|^2 $ is the fluctuation intensity and $\displaystyle L_I^{-1}=\frac{1}{I_k} \frac{\partial I_k }{\partial r} $ is the intensity gradient scale length, 
$A_{k^{\prime},k^{\prime\prime}}=\frac{1}{2} \left[ k_y^{\prime 2} \left(k_{\perp}^{\prime\prime 2} - k_{\perp}^{\prime 2} + 2k_x^{\prime\prime 2}\right) + k_y^{\prime\prime 2} \left(k_{\perp}^{\prime2} - k_{\perp}^{\prime\prime2} + 2k_x^{\prime2}\right) \right]\rho_s^4$, and $B_{k^{\prime},k^{\prime\prime}}= \Big(k_y^{\prime2} k_x^{\prime\prime2}-k_y^{\prime\prime2} k_x^{\prime2} \Big) \Big(k_{\perp}^{\prime\prime2} - k_{\perp}^{\prime2} \Big)\rho_s^6$. Similarly, the second and the third nonlinear fluxes are written as
\begin{eqnarray}
\Pi_{r,\theta}^{NL,2}&=&\langle \tilde{v}_r  \tilde{n}^{(c)} \tilde{v}_{\theta} \rangle \nonumber\\
&=& n_0 c_s^2 \Re \sum_{k=k^{\prime} + k^{\prime\prime}} -i k_y^{\prime} \rho_s^2  \bigg\langle \phi_k^{(c)*} \phi_{k^{\prime}} \bigg( \frac{\partial}{\partial r} \phi_{k^{\prime\prime}}(t) + ik^{\prime\prime}_x \phi_{k^{\prime\prime}}(t) \bigg) \bigg\rangle \nonumber\\
&=&-\frac{1}{2} n_0 c_s^2 \sum_{k=k^{\prime} + k^{\prime\prime}}\tau_{c}\omega_{ci}\frac{I_{k^{\prime}}I_{k^{\prime\prime}}}{1+k_{\perp}^2\rho_s^2} \left( \frac{1}{2} A_{k^{\prime},k^{\prime\prime}}\frac{\rho_s^2}{L_I^2}+B_{k^{\prime},k^{\prime\prime}} \right),
\end{eqnarray}
and
\begin{eqnarray}
\Pi_{r,\theta}^{NL,3}&=&\langle \tilde{v}_r \tilde{n} \tilde{v}_{\theta}^{(c)} \rangle \nonumber\\
&=& n_0 c_s^2 \Re \sum_{k=k^{\prime} + k^{\prime\prime}} -i k_y^{\prime}\rho_s^2 \bigg\langle \bigg( \frac{\partial}{\partial r} \phi_k^{(c)*}(t) + ik_x \phi_k^{(c)*}(t) \bigg)\phi_{k^{\prime}} n_{k^{\prime\prime}} \bigg\rangle \nonumber\\
&=&- \frac{1}{2} n_0 c_s^2 \sum_{k=k^{\prime} + k^{\prime\prime}}\tau_{c}\omega_{ci} \frac{I_{k^{\prime}}I_{k^{\prime\prime}}}{1+k_{\perp}^2\rho_s^2} \left( \frac{1}{2} A_{k^{\prime},k^{\prime\prime}}\frac{\rho_s^2}{L_I^2}+B_{k^{\prime},k^{\prime\prime}} \right),
\end{eqnarray}
respectively. Then, the total nonlinear poloidal momentum flux, Eq.~(\ref{NLflux}), can be obtained by taking summation of the above three nonlinear fluxes
\begin{eqnarray}
\Pi^{NL}_{r,\theta}=-\frac{1}{6} n_0 c_s^2 \sum_{k=k^{\prime} + k^{\prime\prime}}\tau_{c}\omega_{ci}\frac{I_{k^{\prime}}I_{k^{\prime\prime}}}{1+k_{\perp}^2\rho_s^2}B_{k^{\prime},k^{\prime\prime}}.
\end{eqnarray}
This is the final expression for nonlinear poloidal momentum flux. We note that the terms related to turbulence intensity gradient cancel with each other. It is seen that the symmetry breaking in fluctuation spectrum is not required for non-zero nonlinear poloidal momentum flux, which is fundamentally different from the case of the Reynolds stress.

\section{III. Comparison of poloidal rotation drive by nonlinear poloidal momentum flux and quasilinear Reynolds stress}

To illustrate the significance of nonlinear poloidal momentum flux, we compare it with the quasiliear Reynolds stress. According to \cite{Diamond book}, for strong turbulence limit, the nonlinear damping rate is estimated as $\gamma^{NL}_k\sim \displaystyle \frac{k_{\perp}^3 \rho_s^3}{1+k_{\perp}^2\rho_s^2} |k_y| c_s I_k^{1/2} $. It follows that $\tau_c \sim \left(\gamma_k^{NL} \right)^{-1}$, and so gives nonlinear poloidal momentum flux $\Pi^{NL}_{r,\theta} \sim -\displaystyle \frac{1}{6} n_0 c_s^2 \frac{B_{k^{\prime},k^{\prime\prime}}}{k_{\perp}^3 \rho_s^3 |k_y| \rho_s}  I_k^{3/2}$. As for the Reynolds stress, \cite{Diamond PoF 1991} showed that the divergence of Reynolds stress is equivalent to the turbulent radial current driven Lorentz force, and the radial current is calculated using quasilinear theory, $J_r \sim n_0 e c_s k_y\rho_s \frac{L_n}{L_s}I_k$. For convenience of comparison, we rewrite both nonlinear and quasilinear poloidal rotation drive in terms of force density. So, the quasilinear Reynolds force density driven by Lorentz force can be written as
\begin{equation}\label{FQL}
F^{QL} \sim - J_r B_t /c \sim - m_i n_0 k_y\rho_s \frac{c_s^2}{\rho_s} \frac{L_n}{L_s} I_k.
\end{equation}
The nonlinear force density is obtained from the negative radial gradient of nonlinear poloidal momentum flux
\begin{equation}\label{FNL}
F^{NL} \sim \frac{1}{6} m_i \frac{\partial}{\partial r} \left( n_0 c_s^2 \frac{B_{k^{\prime},k^{\prime\prime}}}{k_{\perp}^3 \rho_s^3 |k_y| \rho_s} I_k^{3/2} \right).
\end{equation}
Note that nonlinear force requires radial variation of the nonlinear poloidal momentum flux, although a non-zero nonlinear flux does not require any radial variation of turbulence intensity. The sign of the quasilinear force obtained in \cite{Diamond PoF 1991} is negative. The sign of the nonlinear force depends on the profile of nonlinear flux. A positive (negative) gradient corresponds to positive (negative) nonlinear force, which is against (additive to) the quasilinear force.

We use the standard mixing length estimate for fluctuation intensity $I_k \sim \frac{\rho_s^2}{L_n^2} \frac{L_s}{L_n}$ with $L_s$ being the magnetic shear scale length \cite{Diamond PoF 1991}. Other typical parameters are: $n_0\sim 10^{19}m^{-3}$, $ B_t \sim 1T$, $T_i\sim T_e\sim 100eV$, $L_n\sim0.05m$, $q\sim 3$, $R \sim 1.5 m$, $L_s \sim q R$, $ k_{\perp}\rho_s \sim 1 $.
For nonlinear force density, one needs to make an estimate for the radial scale length of nonlinear flux, which is denoted by $L_{\Pi}$.
We take mesoscale, i.e., $L_{\Pi} \sim \sqrt{L_n \rho_s}$. It is useful to estimate the predicted poloidal flow velocity, $V_{\theta}$ as well. Neoclassical magnetic pumping drag is assumed to balance the turbulent flow drive due to both quasilinear and nonlinear force, i.e., $V_{\theta} \approx \left( F^{QL} +F^{NL} \right)/(\mu m_i n_0)$, where $\mu = q v_{thi} /R$. Here, the plateau regime neoclassical viscous damping rate was used \cite{Diamond PoF 1991}. The results for comparisons of turbulent flow drive, predicted poloidal flow velocity and corresponding Mach number are listed in Table~\ref{comparison}. Here, as mentioned before, positive (negative) gradient of nonlinear flux corresponds to positive (negative) nonlinear force, which acts as a damping (driving) regarding to the quasilinear Lorentz force. The order of magnitude of the nonlinear driving/damping and corresponding poloidal flow velocity can be comparable to those of quasilinear case. The ratio of nonlinear force density to quasilinear one is
\begin{equation}
\frac{|F^{NL}|}{|F^{QL}|} \sim \frac{1}{6} \frac{\rho_s}{L_{\Pi}} \frac{\rho_s}{L_{n}} \left( \frac{L_s}{L_n} \right)^{3/2}.
\end{equation}
We can see that the nonlinear poloidal rotation drive tends to be important in steep density gradient regions, such as in internal transport barrier (ITB) and edge transport barrier (ETB). Here, the fluctuation intensity was estimated by using the standard mixing length theory for slablike drift wave. Extension to toroidal plasmas, particularly for ITB case, one may need to be careful about the magnetic field structure, i.e., weak magnetic shear effects on mode structure and turbulent rotation drive. Magnetic shear has been found to paly an important role in toroidal intrinsic torque reversal from both experiments \cite{Rice2013} and gyrokinetic simulations \cite{LuZX2015}.

\begin{table}\caption{Comparison of the polodial flow drive and the predicted poloidal flow velocity. \\
}\label{comparison}
\begin{tabular}[b]{cccc}
  \hline \hline \qquad\qquad& Force ($N/m^{3}$) \qquad \qquad  &$ V_{\theta}$ ($km/s$) \qquad \qquad  & Mach number \\
\hline Qusilinear \quad\quad & -65              \quad \qquad& -22                       \quad\quad & -0.23 \\
Nonlinear         \quad\quad & $\pm$ 32             \quad\quad & $\pm$ 11                     \quad\quad & $\pm$ 0.11       \\
 \hline \hline
\end{tabular}
\end{table}

\section{IV. conclusions and discussions}

In this work, we have calculated nonlinear poloidal momentum flux in strong electrostatic turbulence using Hasegawa-Mima drift wave model. EDQNM theory has been used for dealing with the nonlinear coupling term and solving the coherent component of beat mode. We adopt the quasi-Gaussian approximation for closure modelling of the forth order moment. In contrast to the quasilinear Reynolds stress, we find that symmetry breaking in turbulence spectrum is not required for non-zero nonlinear poloidal momentum flux. However, the poloidal rotation drive by the divergence of nonlinear momentum flux requires a radial inhomogeneity of nonlinear poloidal momentum flux. Our theoretical predictions indicate that nonlinear poloidal momentum flux can be significant to poloidal rotation drive in strong turbulence as compared to the quasilinear Reynolds stress \cite{Diamond PoF 1991}, particularly in steep density gradient regions.

Experimental observations on ASDEX-U indicate that nonlinear poloidal momentum flux is dominated by the ELM burst \cite{Muller NF 2011}. The anomalous poloidal rotation driven by Reynolds stress and its connection to ITB formation was studied in \cite{Tala2007}. However, there are no simulation codes retained the nonlinear poloidal momentum flux and compared to the usual Reynolds stress. According to our theoretical results, in steep density gradient regions, it may be needed to take into account of nonlinear momentum flux for comprehensive understanding poloidal rotation and its effects on transport barrier formation. For strong burst phenomena, such as blobs in the L mode and ELMs in H mode plasmas, nonlinear flux may be important. Therefore, it is also interesting to investigate how statistical feature of blobs affect the nonlinear flux and the nonlinearly driven flow effects on subsequent burst events.


We note that poloidal rotation contribution to $E \times B$ flow shear can suppress the fluctuation level and the associated turbulent drive for nonlinear poloidal momentum flux. Therefore, our future work will focus on self-consistent calculation of nonlinear poloidal rotation drive and its effects on turbulence suppression. Finally, as we mentioned in the end of our previous work \cite{Wang2015}, investigation on the higher order contributions to poloidal momentum flux for weak turbulence may be worthwhile as well. Hence, we also plan to calculate the nonlinear resonant poloidal momentum transport, and study its effects on poloidal rotation drive.

\section*{Acknowledgments}
We are grateful to T. S. Hahm, Q. M. Hu, Z. F. Cheng, Y. Kosuga, Z. B. Guo and the participants in the Festival of Theory, Aix en Provence 2015 for fruitful discussions. This work was supported by the MOST of China under Contract No.~2013GB112002, the NSFC Grant No.~11305071, and U.S. DOE Contract No. DE-FG02-04ER54738.

\appendix
\section{Calculation of nonlinear poloidal momentum flux}
The detailed process of calculation is given in this appendix for the interested readers. Substituting the coherent component $\phi_k^{(c)}$, Eq.~(\ref{coherent component}), into Eq.~(\ref{NLflux}), with help of Eqs.~(\ref{quadratic moments}) and (\ref{one-time correlation}), and we can calculate the three nonlinear fluxes one by one. The first one can be written as
\begin{eqnarray}
\Pi_{r,\theta}^{NL,1}&=&\langle \tilde{v}_r^{(c)} \tilde{n} \tilde{v}_{\theta} \rangle \nonumber\\
 &=& n_0 c_s^2 \Re \sum_{k=k^{\prime} + k^{\prime\prime}} ik_y \rho_s^2 \langle \phi_k^{(c)*} \phi_{k^{\prime}} ( \frac{\partial}{\partial r} \phi_{k^{\prime\prime}} + ik^{\prime\prime}_x \phi_{k^{\prime\prime}} ) \rangle  \nonumber\\
&=& \frac{1}{2} n_0 c_s^2 \Re \sum_{k=k^{\prime} + k^{\prime\prime}} 2 ik_y\rho_s^2 \Bigg[\int_{-\infty} ^t dt^{\prime} exp \left[ \left( -i\omega_k + \gamma^{NL}_k \right) (t^{\prime}-t) \right] M_{k,k^{\prime},k^{\prime\prime}}^*(t^{\prime})\phi_{k^{\prime}} ( \frac{\partial}{\partial r} \phi_{k^{\prime\prime}} + ik^{\prime\prime}_x \phi_{k^{\prime\prime}} ) \nonumber \\
& &+ k^{\prime}\leftrightarrow k^{\prime\prime}\Bigg] \nonumber \\
&=&\frac{1}{2}n_0 c_s^2  \Re \sum_{k=k^{\prime} + k^{\prime\prime}} \frac{i k_y \rho_s^6}{1+k_{\perp}^2\rho_s^2} \Bigg \lbrace \omega_{ci} \int_{-\infty} ^t dt^{\prime} exp \big\lbrace \left[ i \left( -\omega_k +\omega_k^{\prime}+ \omega_k^{\prime\prime} \right) + (\gamma^{NL}_k + \gamma^{NL}_{k^{\prime}} + \gamma^{NL}_{k^{\prime\prime}}) \right] (t^{\prime}-t) \big \rbrace \nonumber \\
&& \times \Bigg[  \hat{z}\times {\bf k}_{\perp}^{\prime} \cdot {\bf k}_{\perp}^{\prime\prime} \bigg(k_{\perp}^{\prime\prime2} - k_{\perp}^{\prime2} \bigg) \bigg\langle \phi_{k^{\prime}}^*(t)\phi_{k^{\prime}}(t)\bigg\rangle   \bigg\langle\phi_{k^{\prime\prime}}^*(t) \bigg( \frac{\partial}{\partial r} \phi_{k^{\prime\prime}}(t) + ik^{\prime\prime}_x \phi_{k^{\prime\prime}}(t) \bigg)   \bigg\rangle \nonumber \\
&&-\displaystyle i \bigg[k_y^{\prime} \Big(k_{\perp}^{\prime\prime2} - k_{\perp}^{\prime2} \Big) - 2k_x^{\prime\prime}\hat{z}\times {\bf k}_{\perp}^{\prime} \cdot {\bf k}_{\perp}^{\prime\prime} \bigg]  \bigg\langle \phi_{k^{\prime}}^*(t)\phi_{k^{\prime}}(t) \bigg\rangle  \bigg\langle \frac{\partial}{\partial r} \phi_{k^{\prime\prime}}^*(t)\bigg( \frac{\partial}{\partial r} \phi_{k^{\prime\prime}}(t) + ik^{\prime\prime}_x \phi_{k^{\prime\prime}}(t) \bigg) \bigg\rangle  \nonumber\\
&&+i \displaystyle \bigg[k_y^{\prime\prime}\Big(k_{\perp}^{\prime\prime2} - k_{\perp}^{\prime2}\Big) - 2k_x^{\prime}\hat{z}\times {\bf k}_{\perp}^{\prime} \cdot {\bf k}_{\perp}^{\prime\prime}\bigg] \bigg\langle \frac{\partial \phi_{k^{\prime}}^*(t)}{\partial r}  \phi_{k^{\prime}}(t) \bigg\rangle \bigg\langle \phi_{k^{\prime\prime}}^*(t)\bigg( \frac{\partial}{\partial r} \phi_{k^{\prime\prime}}(t) + ik^{\prime\prime}_x \phi_{k^{\prime\prime}}(t) \bigg) \bigg\rangle \Bigg] \nonumber\\
&&+ k^{\prime}\leftrightarrow k^{\prime\prime} \Bigg \rbrace \nonumber\\
&=&\frac{1}{2} n_0 c_s^2 \sum_{k=k^{\prime} + k^{\prime\prime}} \frac{ \rho_s^6}{1+k_{\perp}^2\rho_s^2} \Bigg \lbrace \tau_{c}\omega_{ci}\Bigg [ k_y^{\prime 2} \Big(k_{\perp}^{\prime\prime2} - k_{\perp}^{\prime2} + 2k_x^{\prime\prime2}\Big) I_{k^{\prime}} \bigg\langle \frac{\partial \phi_{k^{\prime\prime}}^*(t)}{\partial r} \frac{\partial \phi_{k^{\prime\prime}}(t)}{\partial r}  \bigg\rangle \nonumber \\
&&+ k_y^{\prime2} k_x^{\prime\prime2} \Big(k_{\perp}^{\prime\prime2} - k_{\perp}^{\prime2} \Big)I_{k^{\prime}}I_{k^{\prime\prime}}  +k_y^{\prime\prime2}\Big(k_{\perp}^{\prime2} - k_{\perp}^{\prime\prime2} + 2k_x^{\prime2}\Big)\bigg\langle \frac{\partial \phi_{k^{\prime}}^*(t)}{\partial r} \phi_{k^{\prime}}(t) \bigg\rangle \bigg\langle \phi_{k^{\prime\prime}}^*(t)\frac{\partial \phi_{k^{\prime\prime}}(t)}{\partial r}  \bigg\rangle  \Bigg ]        \nonumber\\
&&+ k^{\prime}\leftrightarrow k^{\prime\prime} \Bigg \rbrace \nonumber\\
&=&\frac{1}{2}n_0 c_s^2 \sum_{k=k^{\prime} + k^{\prime\prime}}\tau_{c}\omega_{ci}\frac{I_{k^{\prime}}I_{k^{\prime\prime}}}{1+k_{\perp}^2\rho_s^2} \left(A_{k^{\prime},k^{\prime\prime}} \frac{\rho_s^2}{L_I^2}+B_{k^{\prime},k^{\prime\prime}} \right).
\end{eqnarray}
Here, the definition of $ I_k $, $L_I$, $A_{k^{\prime},k^{\prime\prime}}$ and $B_{k^{\prime},k^{\prime\prime}}$ have been given in the text before. In the same way, the second nonlinear flux can be written as
\begin{eqnarray}
\Pi_{r,\theta}^{NL,2}&=&\langle \tilde{v}_r \tilde{n}^{(c)} \tilde{v}_{\theta} \rangle \nonumber\\
 &=& n_0 c_s^2 \Re \sum_{k=k^{\prime} + k^{\prime\prime}} -i k_y^{\prime} \rho_s^2 \bigg\langle \phi_k^{(c)*} \phi_{k^{\prime}} \bigg( \frac{\partial}{\partial r} \phi_{k^{\prime\prime}}(t) + ik^{\prime\prime}_x \phi_{k^{\prime\prime}}(t) \bigg) \bigg\rangle \nonumber\\
&=& \frac{1}{2} n_0 c_s^2  \Re \sum_{k=k^{\prime} + k^{\prime\prime}} \Bigg \lbrace\frac{-i k_y^{\prime} \rho_s^6}{1+k_{\perp}^2\rho_s^2} \omega_{ci}\int_{-\infty} ^t dt^{\prime} exp \big\lbrace \left[ i \left( -\omega_k +\omega_k^{\prime}+ \omega_k^{\prime\prime} \right) + (\gamma^{NL}_k + \gamma^{NL}_{k^{\prime}} + \gamma^{NL}_{k^{\prime\prime}}) \right] (t^{\prime}-t) \big \rbrace \nonumber \\
&& \times \Bigg [ \hat{z}\times {\bf k}_{\perp}^{\prime} \cdot {\bf k}_{\perp}^{\prime\prime} \bigg(k_{\perp}^{\prime\prime2} - k_{\perp}^{\prime2} \bigg) \bigg\langle \phi_{k^{\prime}}^*(t)\phi_{k^{\prime}}(t)\bigg\rangle \bigg\langle \phi_{k^{\prime\prime}}^*(t) \bigg( \frac{\partial}{\partial r} \phi_{k^{\prime\prime}}(t) + ik^{\prime\prime}_x \phi_{k^{\prime\prime}}(t) \bigg)   \bigg\rangle \nonumber \\
&&-\displaystyle i \bigg[k_y^{\prime} \Big(k_{\perp}^{\prime\prime2} - k_{\perp}^{\prime2} \Big) - 2k_x^{\prime\prime}\hat{z}\times {\bf k}_{\perp}^{\prime} \cdot {\bf k}_{\perp}^{\prime\prime} \bigg]  \bigg\langle \phi_{k^{\prime}}^*(t)\phi_{k^{\prime}}(t) \bigg\rangle  \bigg\langle \frac{\partial}{\partial r} \phi_{k^{\prime\prime}}^*(t)\bigg( \frac{\partial}{\partial r} \phi_{k^{\prime\prime}}(t) + ik^{\prime\prime}_x \phi_{k^{\prime\prime}}(t) \bigg) \bigg\rangle  \nonumber\\
&&+i \displaystyle \bigg[k_y^{\prime\prime}\Big(k_{\perp}^{\prime\prime2} - k_{\perp}^{\prime2}\Big) - 2k_x^{\prime}\hat{z}\times {\bf k}_{\perp}^{\prime} \cdot {\bf k}_{\perp}^{\prime\prime}\bigg] \bigg\langle \frac{\partial \phi_{k^{\prime}}^*(t)}{\partial r}  \phi_{k^{\prime}}(t) \bigg\rangle \bigg\langle \phi_{k^{\prime\prime}}^*(t)\bigg( \frac{\partial}{\partial r} \phi_{k^{\prime\prime}}(t) + ik^{\prime\prime}_x \phi_{k^{\prime\prime}}(t) \bigg) \bigg\rangle
 \Bigg ]  \nonumber\\
&& + k^{\prime}\leftrightarrow k^{\prime\prime} \Bigg \rbrace \nonumber\\
&=&-\frac{1}{2} n_0 c_s^2 \sum_{k=k^{\prime} + k^{\prime\prime}}\tau_{c}\omega_{ci}\frac{I_{k^{\prime}}I_{k^{\prime\prime}}}{1+k_{\perp}^2\rho_s^2} \times \left( \frac{1}{2}A_{k^{\prime},k^{\prime\prime}} \frac{\rho_s^2}{L_I^2}+ B_{k^{\prime},k^{\prime\prime}} \right).
\end{eqnarray}
The third nonlinear flux can be written as
\begin{eqnarray}
\Pi_{r,\theta}^{NL,3}&=&\langle \tilde{v}_r \tilde{n} \tilde{v}_{\theta}^{(c)} \rangle \nonumber\\
 &=& n_0c_s^2 \Re \sum_{k=k^{\prime} + k^{\prime\prime}} -i k_y^{\prime} \rho_s^2 \bigg\langle \bigg( \frac{\partial}{\partial r} \phi_k^{(c)*}(t) + ik_x \phi_k^{(c)*}(t) \bigg)\phi_{k^{\prime}} n_{k^{\prime\prime}} \bigg\rangle \nonumber\\
&=&\frac{1}{2}n_0 c_s^2  \Re \sum_{k=k^{\prime} + k^{\prime\prime}} \Bigg \lbrace \frac{-ik_y^{\prime} \rho_s^6}{1+k_{\perp}^2\rho_s^2} \omega_{ci}\int_{-\infty} ^t dt^{\prime} exp \big\lbrace \left[ i \left( -\omega_k +\omega_k^{\prime}+ \omega_k^{\prime\prime} \right) + (\gamma^{NL}_k + \gamma^{NL}_{k^{\prime}} + \gamma^{NL}_{k^{\prime\prime}}) \right] (t^{\prime}-t) \big \rbrace \nonumber\\
&&\times \Bigg [ \hat{z}\times {\bf k}_{\perp}^{\prime} \cdot {\bf k}_{\perp}^{\prime\prime} \bigg(k_{\perp}^{\prime\prime2} - k_{\perp}^{\prime2} \bigg) \Bigg\langle \Bigg(\frac{\partial \phi_{k^{\prime}}^*(t)}{\partial r}\phi_{k^{\prime\prime}}^*(t) + \phi_{k^{\prime}}^*(t)\frac{\partial \phi_{k^{\prime\prime}}^*(t)}{\partial r}\Bigg)   \phi_{k^{\prime}}(t) \phi_{k^{\prime\prime}}(t) \Bigg\rangle \nonumber \\
&&-\displaystyle i \bigg[k_y^{\prime} \Big(k_{\perp}^{\prime\prime2} - k_{\perp}^{\prime2} \Big) - 2k_x^{\prime\prime}\hat{z}\times {\bf k}_{\perp}^{\prime} \cdot {\bf k}_{\perp}^{\prime\prime} \bigg]   \bigg\langle \Bigg(\frac{\partial \phi_{k^{\prime}}^*(t)}{\partial r}\frac{\partial \phi_{k^{\prime\prime}}^*(t)}{\partial r} +\phi_{k^{\prime}}^*\frac{\partial^2 \phi_{k^{\prime\prime}}^*}{\partial r^2}\Bigg) \phi_{k^{\prime}}(t) \phi_{k^{\prime\prime}}(t) \bigg\rangle  \nonumber\\
&&+i \displaystyle \bigg[k_y^{\prime\prime}\Big(k_{\perp}^{\prime\prime2} - k_{\perp}^{\prime2}\Big) - 2k_x^{\prime}\hat{z}\times {\bf k}_{\perp}^{\prime} \cdot {\bf k}_{\perp}^{\prime\prime}\bigg] \bigg\langle \Bigg(\frac{\partial \phi_{k^{\prime}}^*(t)}{\partial r}\frac{\partial \phi_{k^{\prime\prime}}^*(t)}{\partial r} +\phi_{k^{\prime\prime}}^*\frac{\partial^2 \phi_{k^{\prime}}^*}{\partial r^2}\Bigg)\phi_{k^{\prime}}(t) \phi_{k^{\prime\prime}}(t) \bigg\rangle \nonumber\\
&&+\hat{z}\times {\bf k}_{\perp}^{\prime} \cdot {\bf k}_{\perp}^{\prime\prime} \bigg(k_{\perp}^{\prime\prime2} - k_{\perp}^{\prime2} \bigg) \Bigg\langle ik_x\phi_{k^{\prime}}^*(t)\phi_{k^{\prime\prime}}^*(t)\phi_{k^{\prime\prime}}^*(t)   \phi_{k^{\prime}}(t) \phi_{k^{\prime\prime}}(t)   \Bigg\rangle \nonumber \\
&&-\displaystyle i \bigg[k_y^{\prime} \Big(k_{\perp}^{\prime\prime2} - k_{\perp}^{\prime2} \Big) - 2k_x^{\prime\prime}\hat{z}\times {\bf k}_{\perp}^{\prime} \cdot {\bf k}_{\perp}^{\prime\prime} \bigg]   \bigg\langle ik_x\phi_{k^{\prime}}^*(t)\frac{\partial \phi_{k^{\prime\prime}}^*(t)}{\partial r}\phi_{k^{\prime}}(t) \phi_{k^{\prime\prime}}(t) \bigg\rangle  \nonumber\\
&&+i \displaystyle \bigg[k_y^{\prime\prime}\Big(k_{\perp}^{\prime\prime2} - k_{\perp}^{\prime2}\Big) - 2k_x^{\prime}\hat{z}\times {\bf k}_{\perp}^{\prime} \cdot {\bf k}_{\perp}^{\prime\prime}\bigg] \bigg\langle \frac{\partial \phi_{k^{\prime}}^*(t)}{\partial r}ik_x\phi_{k^{\prime\prime}}^*(t)\phi_{k^{\prime}}(t) \phi_{k^{\prime\prime}}(t) \bigg\rangle \Bigg ]  \nonumber\\
&& + k^{\prime}\leftrightarrow k^{\prime\prime} \Bigg \rbrace  \nonumber \\
&=&- \frac{1}{2} n_0 c_s^2 \sum_{k=k^{\prime} + k^{\prime\prime}}\tau_{c}\omega_{ci} \frac{I_{k^{\prime}}I_{k^{\prime\prime}}}{1+k_{\perp}^2\rho_s^2}\left( \frac{1}{2} A_{k^{\prime},k^{\prime\prime}} \frac{\rho_s^2}{L_I^2}+ B_{k^{\prime},k^{\prime\prime}}  \right).
\end{eqnarray}
Then, the total nonlinear poloidal momentum flux can be obtained by taking summation of the above three nonlinear fluxes
\begin{eqnarray}
\Pi^{NL}_{\theta}&=&\frac{1}{3} \left( \langle \tilde{v}_r^{(c)} \tilde{n} \tilde{v}_{\theta} \rangle + \langle \tilde{v}_r \tilde{n}^{(c)} \tilde{v}_{\theta} \rangle + \langle \tilde{v}_r \tilde{n} \tilde{v}_{\theta}^{(c)} \rangle \right) \nonumber \\
&=&-\frac{1}{6} n_0 c_s^2 \sum_{k=k^{\prime} + k^{\prime\prime}}\tau_{c}\omega_{ci}\frac{I_{k^{\prime}}I_{k^{\prime\prime}}}{1+k_{\perp}^2\rho_s^2}B_{k^{\prime},k^{\prime\prime}} \nonumber
\end{eqnarray}

\end{document}